# Evolution of an Emerging Symmetric Quantum Cryptographic Algorithm

**Omer K. Jasim[1*], Safia Abbas[2], El-Sayed M. Horbaty[2], Abdel-Badeeh M. Salem[2]**

[1]Computer Science, Al-Ma'arif College University, Anbar, Iraq
[2]Faculty of Computer and Information Sciences, Ain Shams University, Cairo, Egypt
Email: [*]omer.k.jasim@ieee.org, safia_abbas@cis.asu.edu.eg, Shorbaty@cis.asu.edu.eg, absalem@cis.asu.edu.eg

## Abstract

**With the rapid evolution of data exchange in network environments, information security has been the most important process for data storage and communication. In order to provide such information security, the confidentiality, data integrity, and data origin authentication must be verified based on cryptographic encryption algorithms. This paper presents a new emerging trend of modern symmetric encryption algorithm by development of the advanced encryption standard (AES) algorithm. The new development focuses on the integration between Quantum Key Distribution (QKD) and an enhanced version of AES. A new quantum symmetric encryption algorithm, which is abbreviated as Quantum-AES (QAES), is the output of such integration. QAES depends on generation of dynamic quantum S-Boxes (DQS-Boxes) based quantum cipher key, instead of the ordinary used static S-Boxes. Furthermore, QAES exploits the specific selected secret key generated from the QKD cipher using two different modes (online and off-line).**

## 1. Introduction

Recently, two distinct ciphers for cryptography processes are defined, symmetric and asymmetric cipher [1]. Symmetric is a form of cryptosystem in which encryption and decryption processes are preformed based on the same key. While, asymmetric encryption and decryption processes are depending mainly on performance using the keys mechanism (public and private) [2]. A Key is a numeric or may be a special symbol which is used through the time of encryption and decryption. The selection of key in cryptography is very important since the security of encryption algorithm depends directly on it. The strength of the encryption algorithm relies on the secrecy of the key, length of the key, the initialization vector, and how they all work together [3].

[*]Corresponding author.

Asymmetric encryption techniques are about 1000 times slower than symmetric encryption which makes it impractical when trying to encrypt large amounts of data [4]. Also to get the same security strength as symmetric, asymmetric must use a stronger key than symmetric encryption technique [5]. Therefore, in this paper we focus on the symmetric cipher technique.

Generally, symmetric technique divided into two broad categories: stream cipher and block cipher [2] [3]. In a stream cipher, an encryption/decryption process embraces on a symbol by symbol at a time. While, a block cipher divides a plaintext into a number of equal blocks (b > 1), these blocks are encrypted together to produce a cipher text [4] [5].

Moreover, a block cipher transforms 64; 128; 192 or 256-bit to string of the same length under the control of secret key [6]. All block cipher systems rely on substitution-permutation boxes, which are fixed and do not have any relation with a secret key. It is adding more complexity to the cipher text by changing the location of the plain text character into a new position [7].

This paper discusses the AES block cipher symmetric algorithm. Scalability, easy to implement and resistance against attacks are well-known features for such algorithm [6] [8] [9]. AES algorithm utilizes the same key for encryption/decryption process with length 128; 192; 256-bits. This key is grouped (XOR-gate) with separate equal blocks of a plain text [10]. Additionally, the length of secret key determines the number of rounds in such algorithm (Nr). Thus, AES-128 bits needs 10 rounds, 12 rounds for AES-192 bits and 14 rounds for AES-256 bits; for more details see [6]. **Table 1** shows the relation among key length (Nk), input block (Nb) and number of rounds (Nr) in the AES algorithm.

Unfortunately, most of modern encryption algorithms rely entirely on a complex mathematics computation in the key generation and distribution and management. So, it is vulnerable to quantum attack and the man-in-the-middle attack. Quantum computation especially QKD is the first solution to these problems.

QKD is an important practical application of quantum computation. It is based on laws of physics rather than computation complexity of mathematical problems. It can generate security keys between two communications and guarantee security of sensitive data (for more details see [11] [12]).

This paper discusses and analyzes the structure of Substitutions boxes (S-boxes) in the traditional AES. Moreover, based on a quantum cipher key, a dynamic quantum S-Boxes (DQS-Boxes) is generated. Finally, a new version of AES by integration between the enhanced version of AES and QKD in two implemented modes (on-line and off-line) is presented.

The rest of the paper is organized as follows: Section 2 surveys the existing study for AES optimization and cryptanalysis. The developed architecture for QAES and integration modes are given in Section 3. Section 4 explains the inferences obtained from the results and discussion. Section 5 presents the conclusion and future works.

## 2. Existing Methods

Recently, there are few authors are focused on the enhancing and cryptanalysis of the AES algorithm, because most of AES attacks appeared lately.

For example, Sekar *et al.* [2] propose a new innovative method to enhance the AES algorithm by increasing the key length to 512 bits and thereby the number of rounds is increased in order to provide a stronger encryption method for secure communication. This method doesn't modify the structure of AES, but only increase the number of rounds. Therefore, this algorithm increases the processing time which will limit the use of AES in real applications.

Kazys *et al.* [13] present a new version of AES by generating random S-Boxes coinciding with every secret key generation. The authors described in details how to generate random S-Box, key-independent, and ratio of independence for the S-Box elements is computed. The breach of this study was not debating any type of cryptanalysis attacks.

**Table 1.** AES key, input block, and rounds.

| Cipher | Nk-words | Nb-words | Nr |
|---|---|---|---|
| AES-128 | 4 | 4 | 10 |
| AES-192 | 6 | 4 | 12 |
| AES-256 | 8 | 4 | 14 |

Shaaban *et al.* [14] develop a powerful algorithm for cryptography, it's based on AES to generate different sub keys from the original key and using each of one to encrypt a single AES round. It becomes resistant against the analysis attacks such as a brute force attack and others. Moreover, the authors classified secret keys as real key and PRNG, each one used in special cryptographic mode. However, this algorithm is slower than classical AES, therefore, it is apt to the timing attack.

However, contrast to above studies, the first cryptanalysis deployed by Alex B. *et al.* [7] [15] when they evaluate the cost of cryptanalytic attacks on the full AES by using a special-purpose hardware in the form of multi-core AES processors. Also, they analyze different time-cost trade-offs and evaluate the implications of progress in VLSI technology under the assumption that Moore's law will continue to hold for the next ten years. These calculations raise some concerns about the long-term security of the AES.

S. Hadi *et al.* [9] propose a new related-key impossible differential attack on 7-round AES-128, which is the first attempt using this technique. They attacked 7-round AES-128 with the time complexity of $(105)^2$, the fastest attack of all the previous ones from time and pre-computation complexities points of views. This attack is the first related key impossible differential attack which is applied to 7-round AES-128. A fundamental point to construct such attack is using a special property of Mix Column operation of AES.

Other studies [12] [16]-[19] discuss the weakness of key schedules (key expansion [6]) which may be used in attacking the block cipher system, until now not have any key schedule attack have been fully broken the AES algorithm.

Finally, we can conclude that many studies have attempted to enhance the strength and efficiency of AES algorithm without changed in a core of AES structure. In other hand, not fully rounds hacked by the attackers, the best states reached to seven rounds (out of ten) for AES-128, and ten (out of 12 or 14) rounds for AES-192 and AES-256.

## 3. QAES Architecture

This section illustrates the QAES development steps based on two different modes (on-line and off-line modes). The main machine utilizes the Core i5 (4.8 GHz) with 8 GB of RAM with 500 GB-HDD, while, the simulator is programmed using Visual Studio Ultimate 2012 (VC#).

### 3.1. QAES Development

The QAES developed system incorporates both the QKD and an enhanced version of the AES in order to provide an unconditional security level [20] [21] for any cipher system built on symmetric encryption algorithms. As shown in **Figure 1**, the AES enhanced version utilizes the dynamic quantum S-Box (DQS-Boxes) that is generated from the QKD and exploits the generated key in the encryption/decryption process.

The DQS-boxes enjoy the dynamic mechanism, in which the contents of each S-Box changes consequently in each round with the change of the key generation. Such dynamic mechanism aids in solving the mechanism problems associated with the traditional S-Boxes. Avoiding the off-line analysis attack, overcoming the DS-Box [14], and resistance to the quantum attack are examples of the associated problems.

Since the unconditional security depends on the Heisenberg uncertainty principle (core feature in quantum theory) [20] [21], instead of the complex mathematical model in key generation, more attack resistance is assured and the cipher system is hard to be attacked.

### 3.2. Integration Modes

The integration between the enhanced AES and the QKD uses two different modes the online and the off-line is explained in this section.

#### 3.2.1. Online Mode

During the negotiation between the two parties (master, slave), as shown in **Figure 2**, the encryption/decryption process is achieved coinciding with quantum key generation.

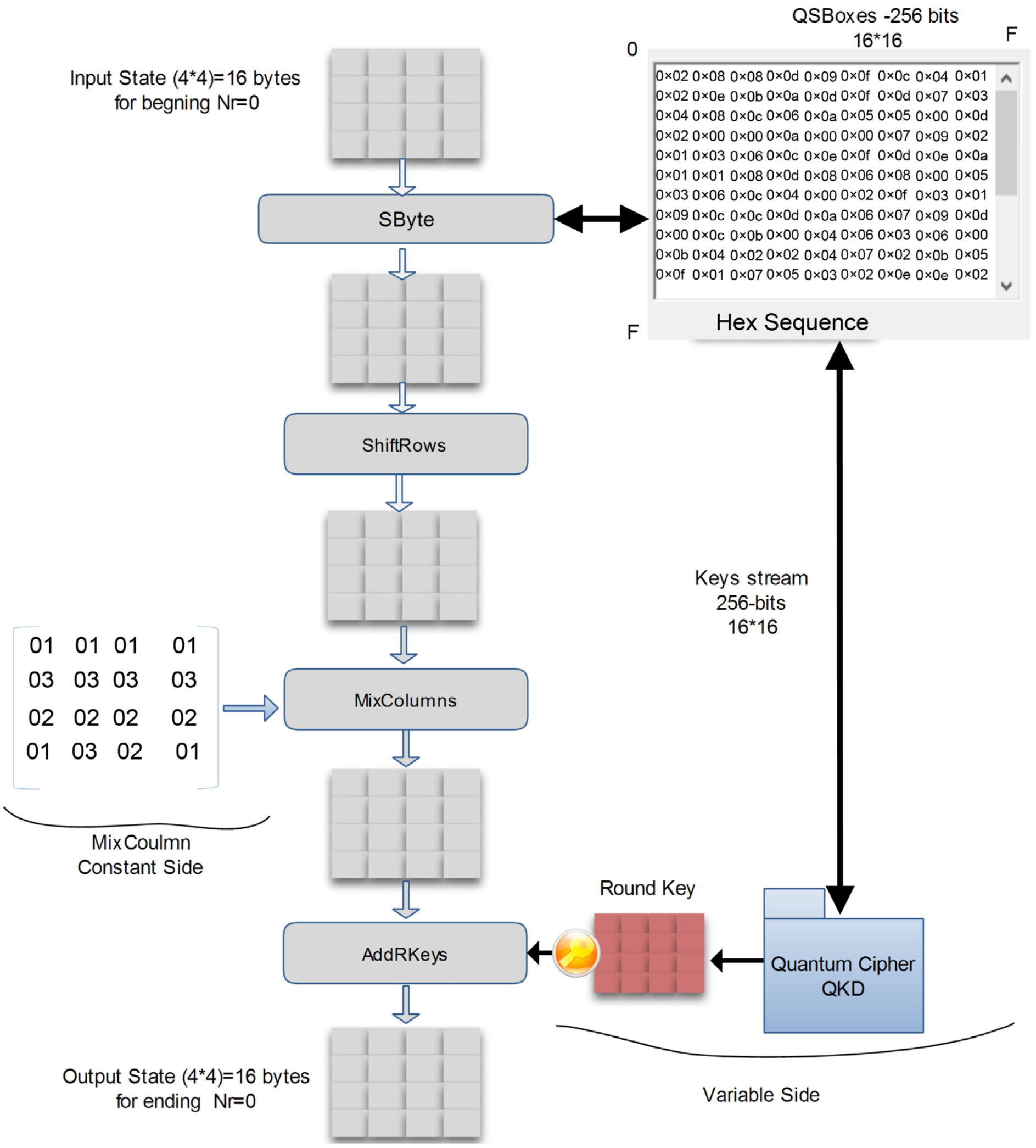

**Figure 1.** QAES architecture for single cycle.

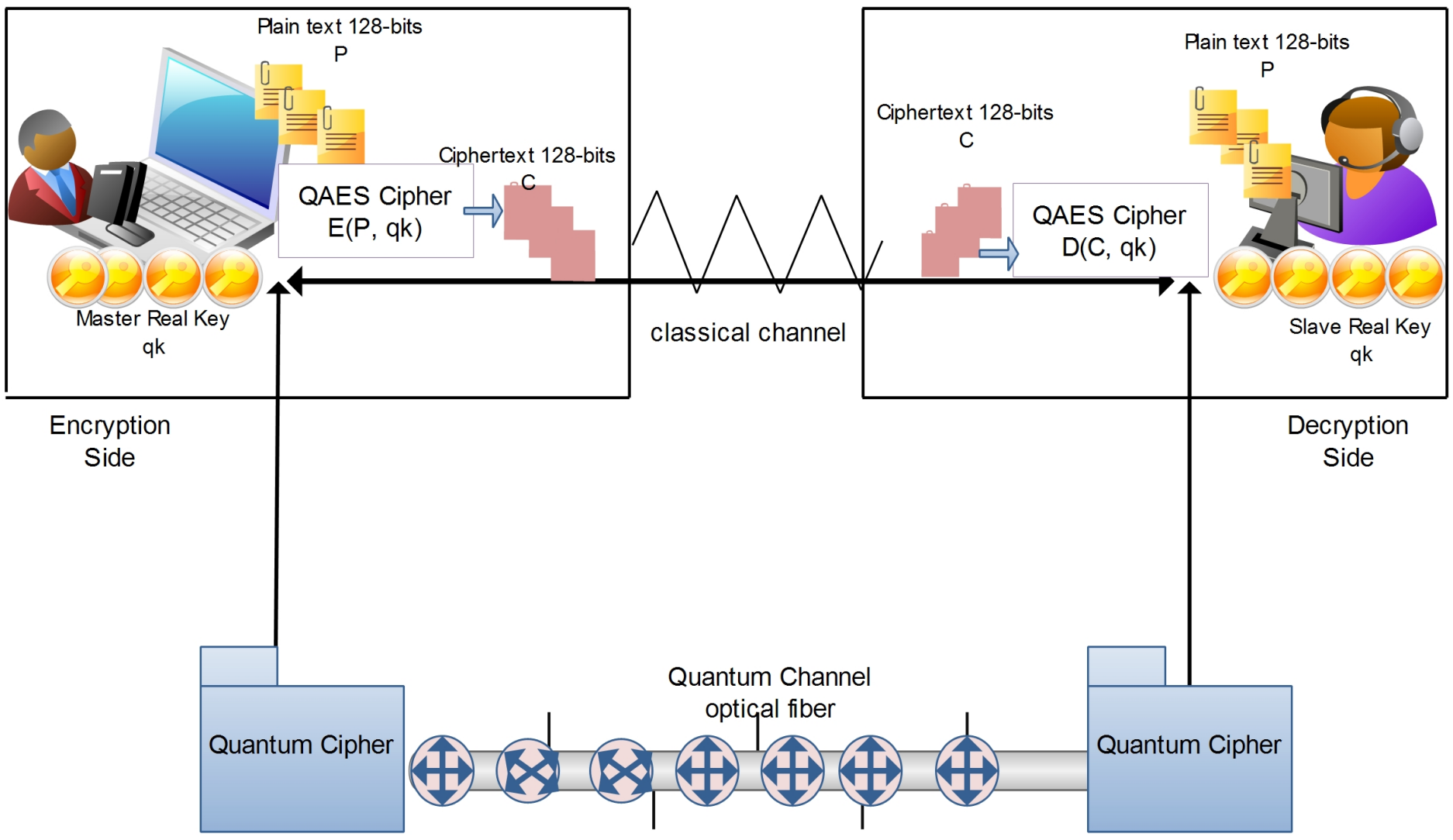

**Figure 2.** On-line mode architecture.

The on-line mode follows the following steps:

- The quantum secret key is generated over the quantum channel using BB84 protocol.
- The master and slave parties check the online compatibility for the generated secret key.
- The master and the slave choose the appropriate key length (128; 192; 256 bits) through the classical channels in order to perform the encryption/decryption process.
- The two parts deploy the selected final quantum secret key ($q_k$) to the symmetric encryption algorithm (AES).
- The system automatically creates the DQS-Box based on the secret key generation with length 256-bits.
- Encrypt the first block input file (P1-128 bits) by the AES stages—using $qk_1$ which generate by QKD round1.

$$E\left(P_1 \oplus qk_1\right) = C_1$$

- Finally, Encrypt final block input file by the AES stages—using $qk_n$ which generate by QKD round$_n$, where $n = Nr = 10$; 12; 14.

$$E\left(P_n \oplus qk_1\right) = C_n$$

- The decryption process start with the end of the encryption process (inverse methodology).

$$D\left(C_n \oplus qk_n\right) = P_n$$

Due to the key availability (KA) associated with QKDs [21] [22], the online mode is provided by a sequence of unrelated keys $\left(qk_1, qk_2, \cdots, qk_n\right)$ in each round, illustrated in **Figure 3**, these unrelated keys prevent the attackers from detecting the next key generation. Then, each QAES round, will consider the generated sequence of keys $\left(qk_1, qk_2, \cdots, qk_n\right)$ as a sequence of sub keys, which in turn are used in the encryption/decryption process.

Finally, this mode can be used with any type of encryption modes such as cipher feedback (CFB) mode, output feedback (OFB) mode, and the counter (CTR) mode.

### 3.2.2. Offline Mode

Despite of many positive criteria for QKD such as KA, key distribution and management, they depend on exchanging photons between parties over limited distances (314 km) [23] [24]. Various randomness tests for qubits generation from QKD based on NIST, DIEHARD, and others were implemented previously [21] and reveals an astonishing fact, that all tests generate truly random bits, that used a key for the encryption/decryption process, in both suggested modes.

As shown in **Figure 4**, the architecture of this mode consists of quantum device, key controller and two communication parties (master-slave). The quantum cipher (QC) used to generate sequence qubits between parties. The QC responsible of exchanging and replacement of keys in both parties (master-slave) and system (user-system).

The off-line mode follows the following steps:

- Generate the random qubits from QKD, these qubits act as secret key (k).
- Based on this K, the system creates DQS-Box.
- Using classical AES-key scheduler, the sub-keys for each AES-round are generated.
- Finally, steps (4, 5, 6, 7 and 8) in the on-line mode are typically performed.

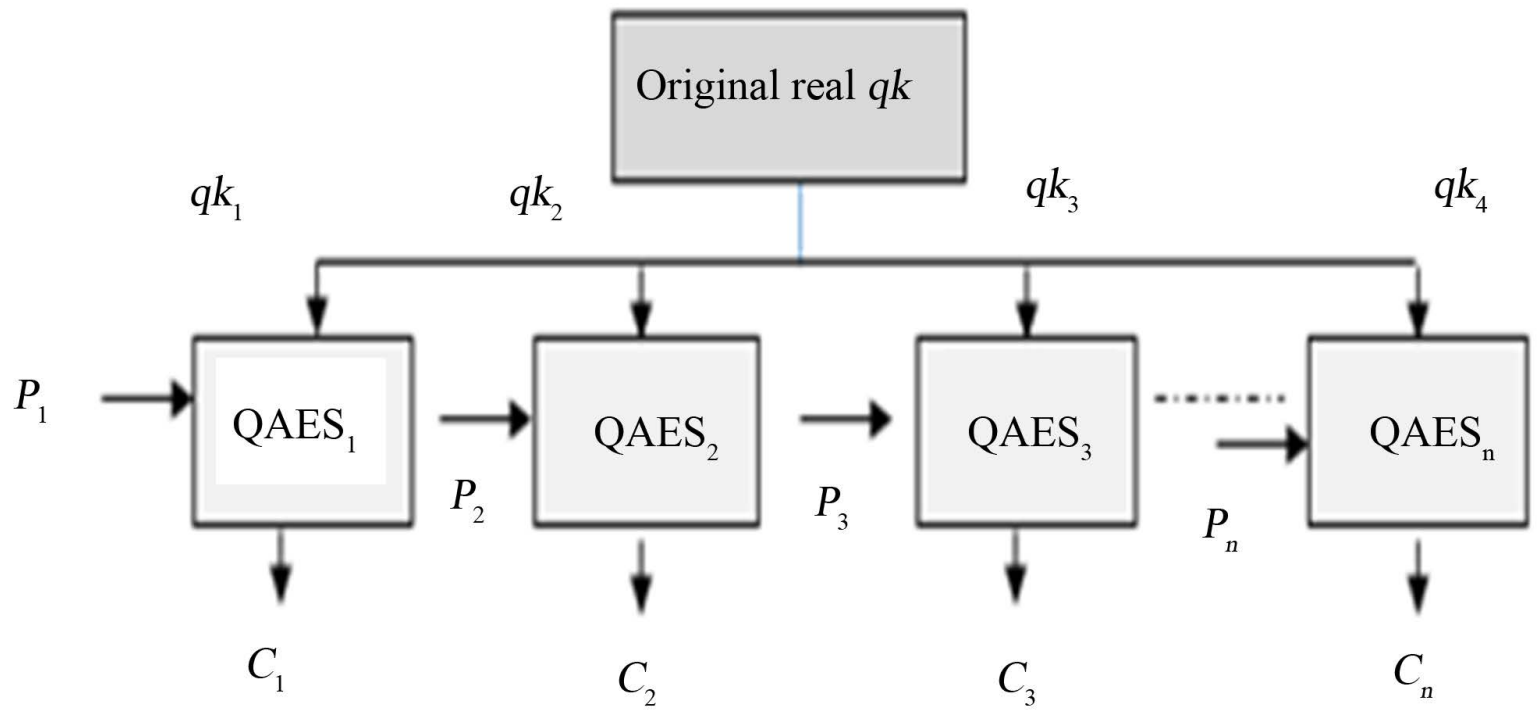

**Figure 3.** On-line encryption mode.

Due to the results of the random tests implementation mention in [21], the unrelated key generation offers a highly secured cipher system. The differences between the QAES and the classical AES are the DQS-Boxes and the generation method of the sub key. As shown in **Figure 5**, this mode can be used with any type of encryption such as CFB, OFB, and CTR.

## 4. Results and Discussions

In this section the ratio of independency for DQS-Box and the time of encryption process have been implemented, measured and analyzed based QAES. After then, the results are compared when the same processes are implemented based classical AES.

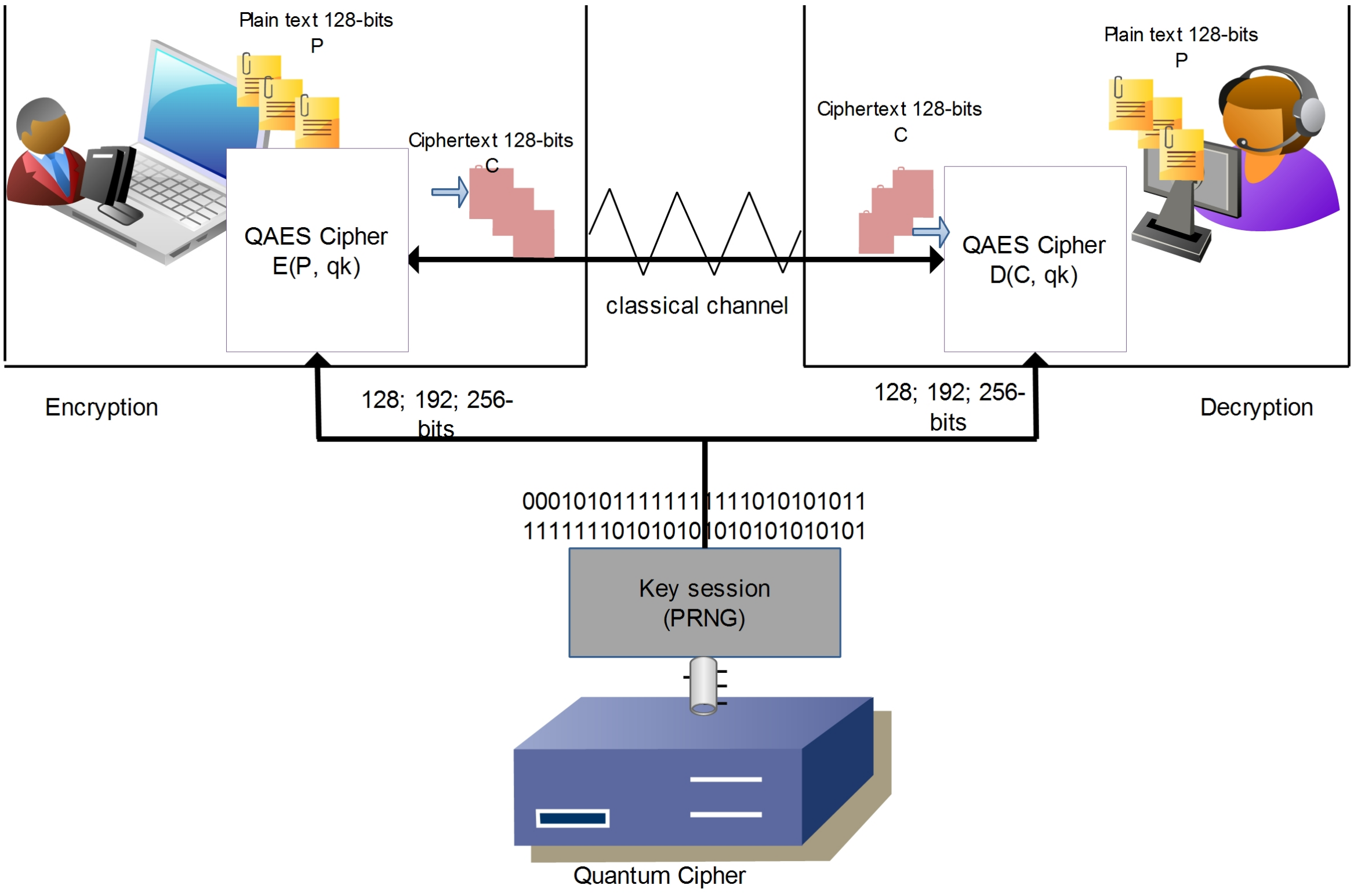

**Figure 4.** Off-line mode.

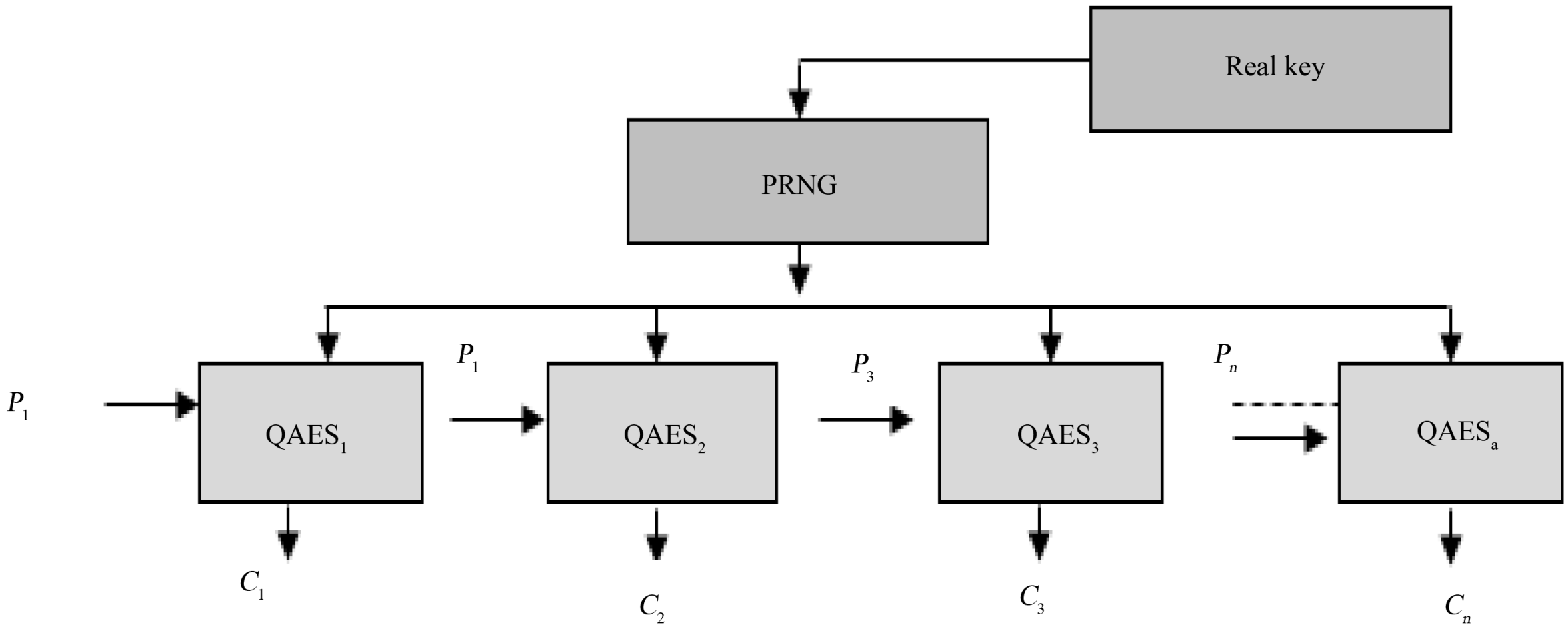

**Figure 5.** Off-line encryption mode.

### 4.1. DQS-Box Analysis

The ratio of independent mainly depends on Equations (1)-(4), these equations are used for the classical S-boxes correlation and independency computations [2] [9]. Accordingly, the ratio of independence based the DQS-Box are calculated. The results show that using the DQS-Box produces higher ratio and lower correlation coefficient between different rounds. This guarantees a better security and more resistance against the almost of cryptanalysis attacks and build a stronger encryption algorithm.

$$y = \frac{x - \text{mean}(x)}{std(x)} \tag{1}$$

$$\text{mean}(x) = \frac{\max(x) - \min(x)}{2} \tag{2}$$

$$std(x) = \sqrt{\max(x) - \min(x)^2 / 16} \tag{3}$$

$$\text{ratio} = \frac{std(CORR)}{\max(x)} \tag{4}$$

where *CORR* is a ratio of correlations.

For example, regarding to the simulation environment for QKD, two DQS-Boxes are generated with 256 bits (matrix 16 × 16). According to Equations (1)-(4), the ratio of correlation functions and independence between DQS-Box1 and DQS-Box2 are shown in **Figure 6**, **Figure 7**.

As shown in the above figures, the ratio of correlation (CORR) coefficient is 6.606%, while the ratio of independence is 93.359%. Exactly, the ratio of independent ranging between (72.78% - 100%) for each corresponding row between two DQS-Boxes, [for more details see **Appendix**], which indicates that the QAES provides a highly unrelated Box that led to a more secured connection. Moreover, the DQS-Box generation process enjoys the advantage that the intermediate generation done during the encryption/decryption process which avoids the offline analysis attack associated with the S-Box.

### 4.2. Efficiency of QAES

In the following analysis, both the AES and QAES algorithms have been implemented using several input files sizes: 500 kb, 1000 kb, 1500 kb, 2000 kb, and 3500 kb.

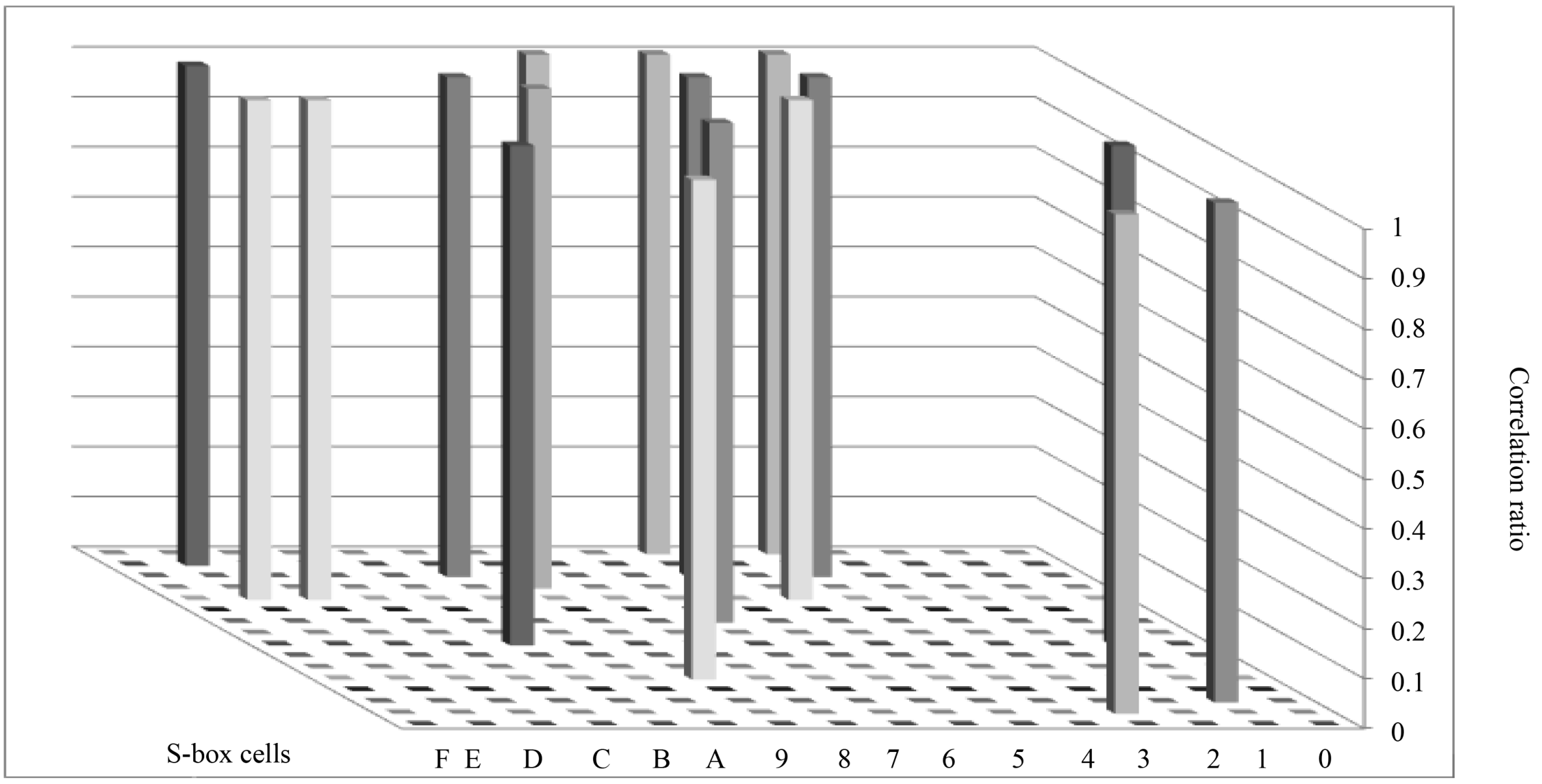

**Figure 6.** Correlation between QS-Box1 and QS-Box2.

**Figure 8** and **Figure 9** represent the running time of the implemented algorithms using the local machines described above, the running time is calculated in milliseconds (ms) and the input size is taken in kilobytes.

Comparing the QAES with traditional AES encryption algorithms reflects a higher security level. However, as shown in Equation (5), and **Figure 10**, this algorithm takes time more than others due to the time required for quantum key generation (time for key negotiation ($T_{QKG}$) and time required for the encryption/decryption process $T(Enc(P))$.

$$T_{qencryption} = T_{QKG} + T\left(E\left(P_n\right)\right) \quad (5)$$

where $T_{qenc}$ = Total encryption based QAES and $P_n$ = plain file.

As showed in **Figure 10**, any change in the eavesdropper activity or noise level directly impact on the length of the secret key and the time of generation it. For example, in order to get 200 from 500 qubits are pumped under the influence of noise 0.05 GHz and there is no Eve influence, we need 0.23 milliseconds to generate it. Usually, this time simultaneously grows with the increasing of the noise or Eve influence. However, the practical environment is faster than the simulator environment, due to the light nature [8] [24].

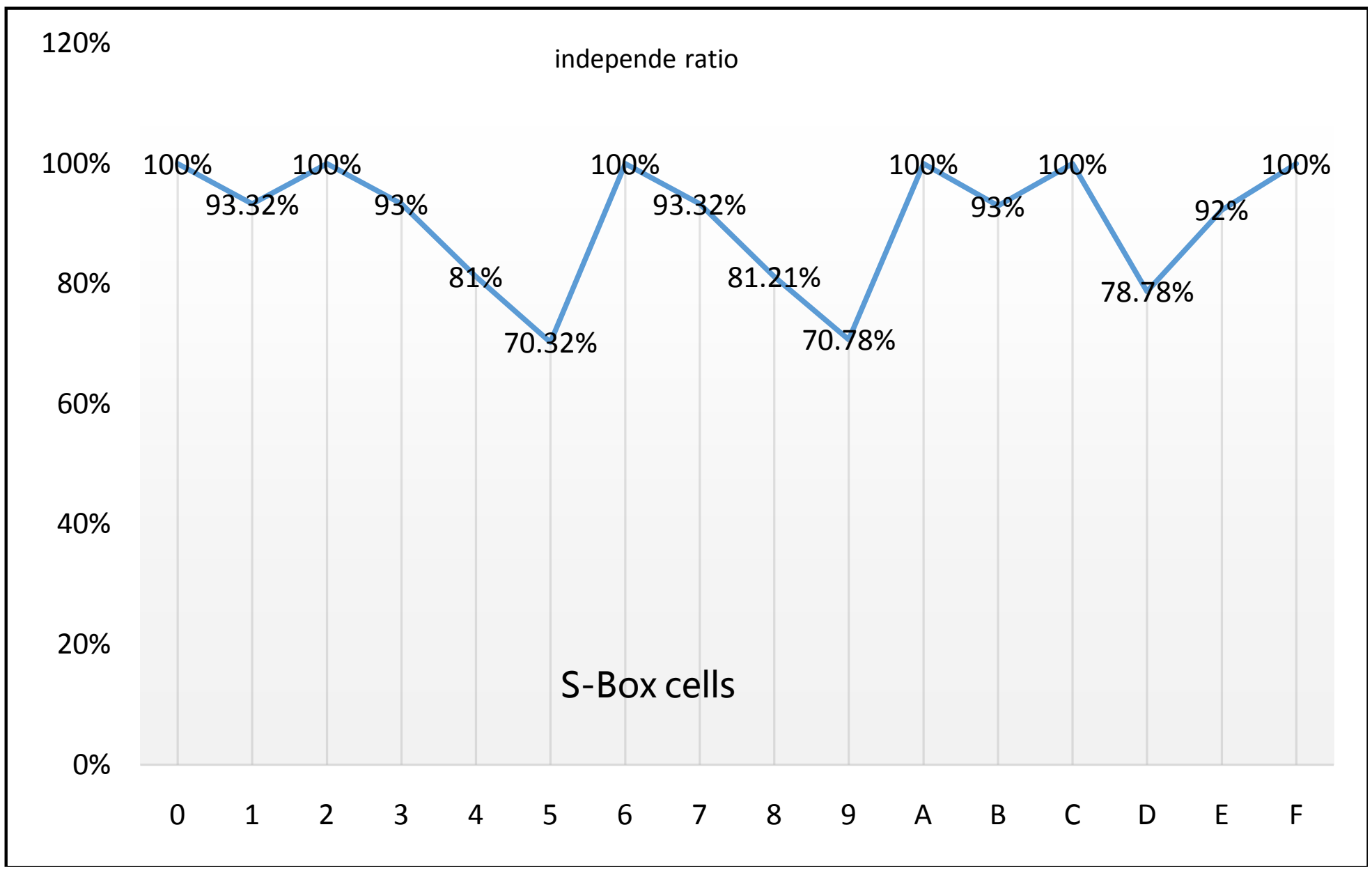

**Figure 7.** Ratio of independence.

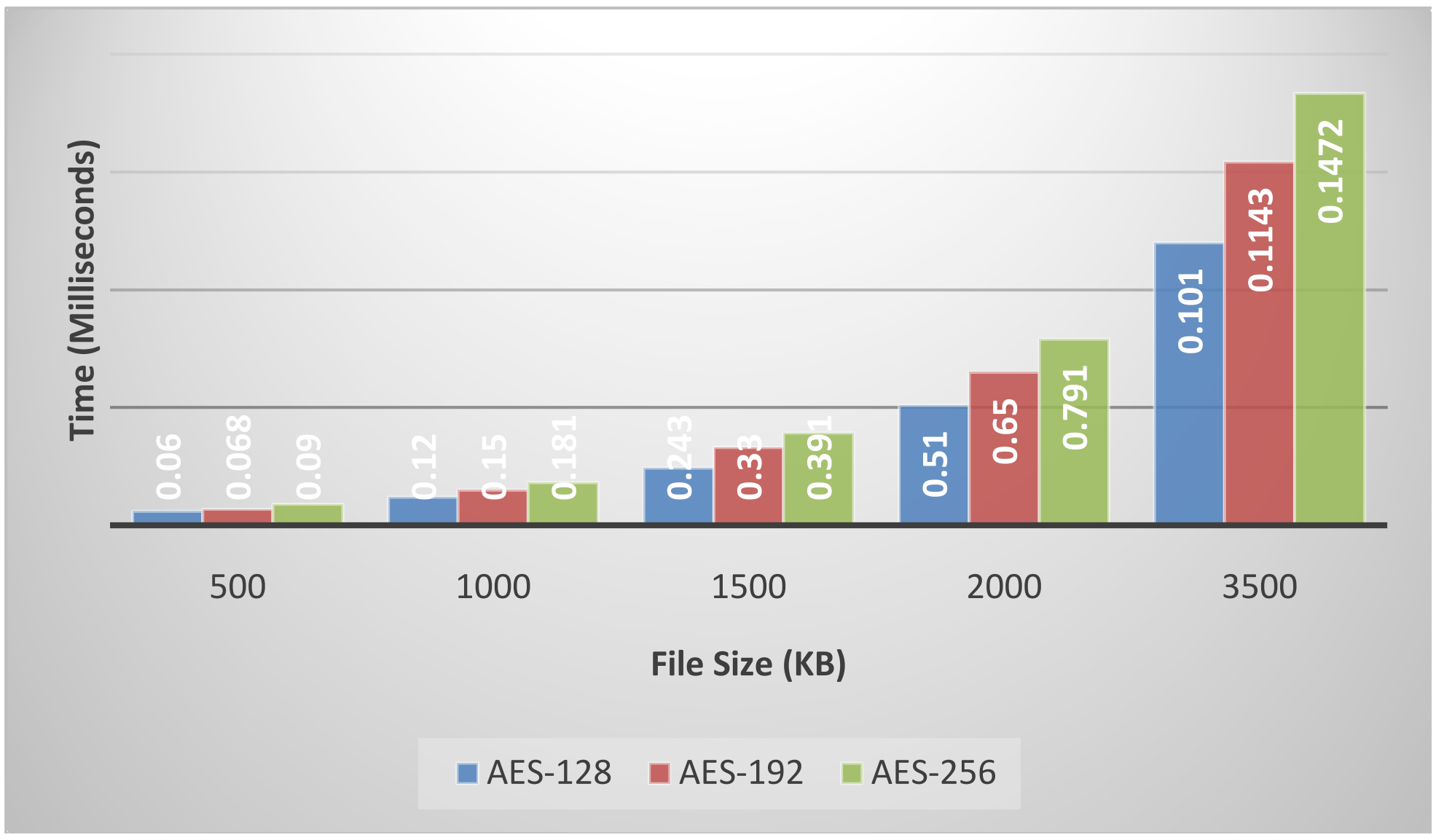

**Figure 8.** AES efficiency.

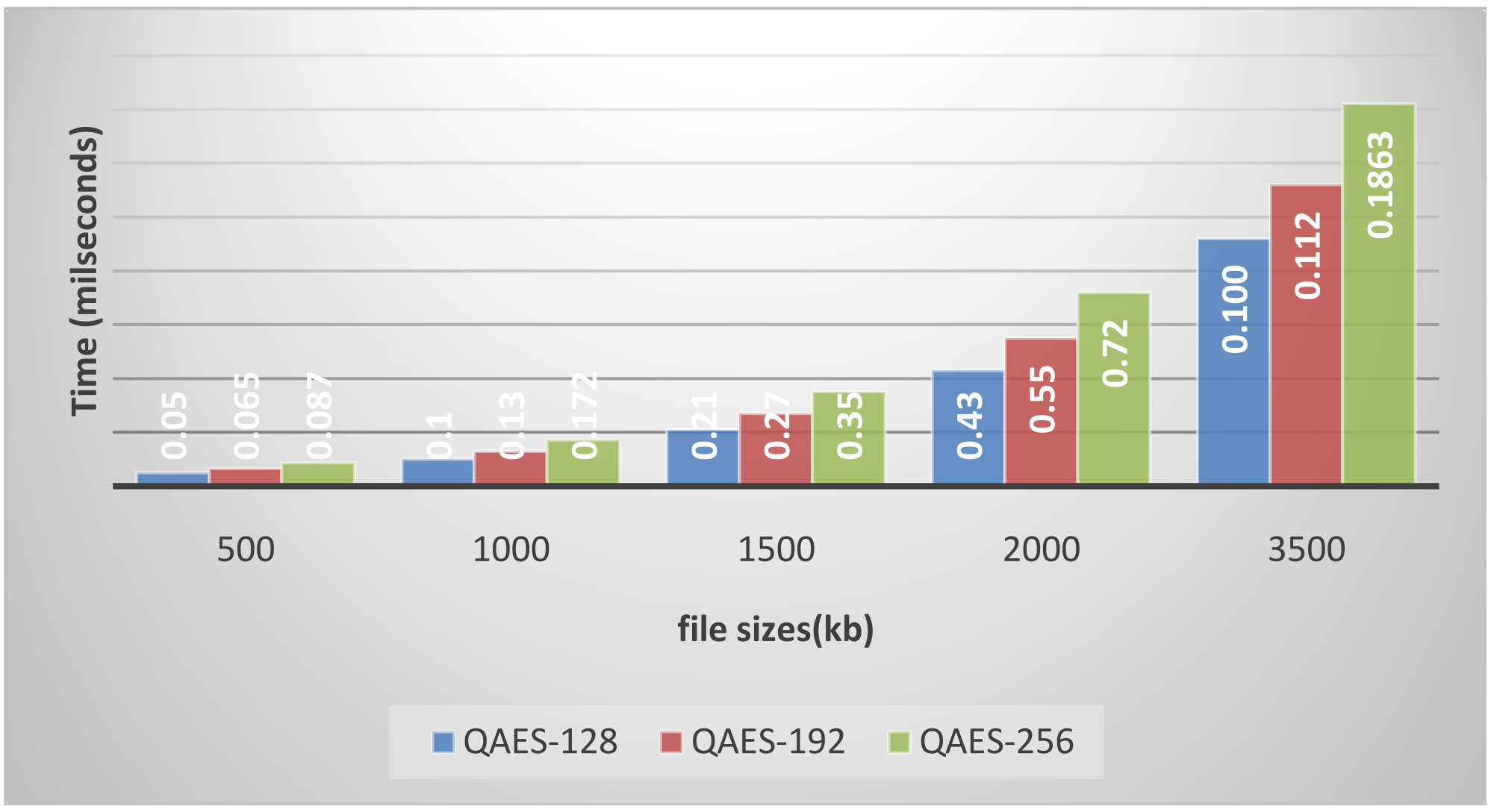

**Figure 9.** QAES efficiency without key generation time.

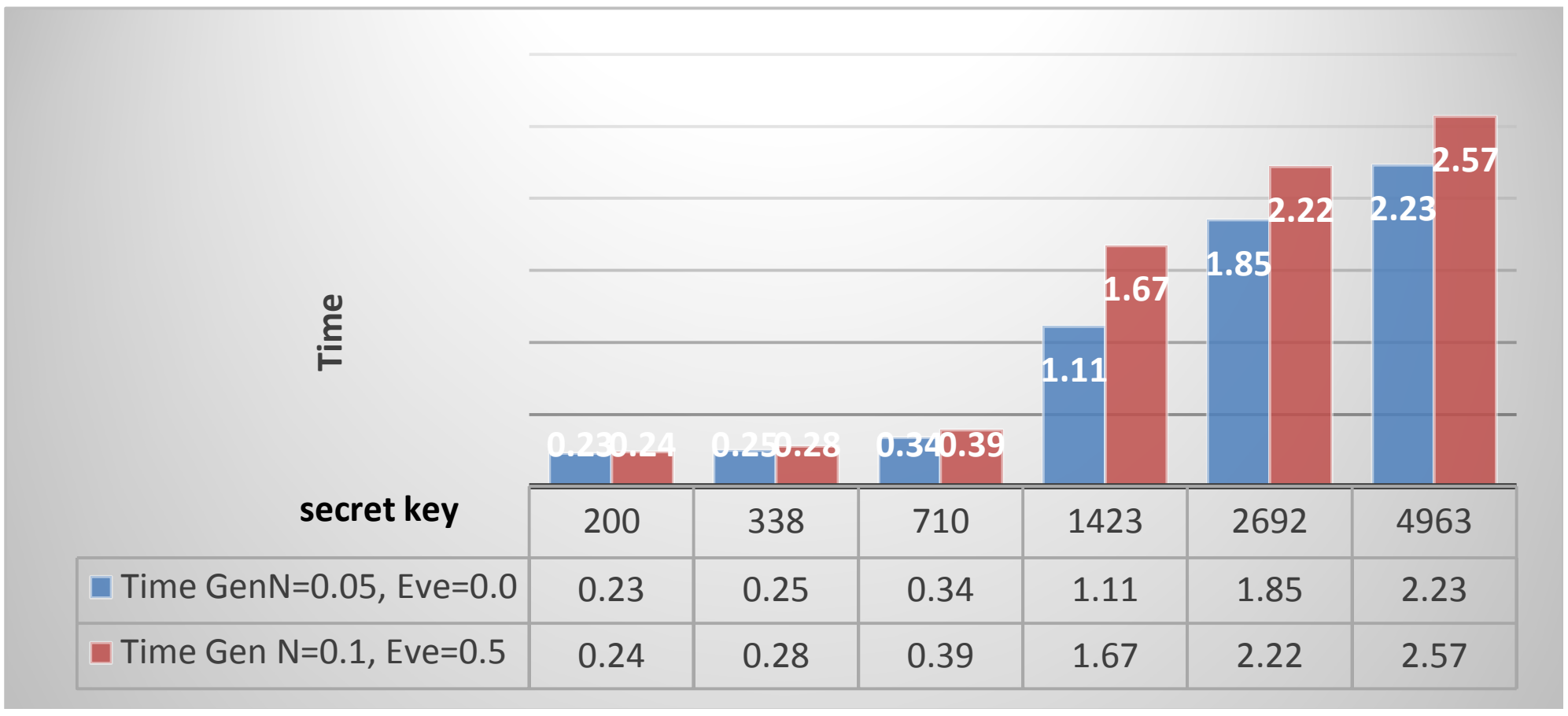

**Figure 10.** Secret key generation with viarouse configurations.

Generally, we can conclude that the QAES is a little bit slower than the AES. For example, according to **Figures 8-10**, and Equation (5), the encryption time of AES-128 bits is 0.101 ms for file size 3500 kb. While, QAES-128 bits takes 0.23 ms for key generation has length 200-qubits and 0.100 ms for file encryption. So, the total time for QAES-128 is 0.123 ms and 0.135 ms for QAES-192.

Finally, since the QAES follows the same architecture of the AES, the input file size has always changed during encryption process and the details of the processed file remain unchangeable.

## 5. Conclusions and Future Works

In this paper a developed version of AES algorithm, annotated as QAES, based on quantum encryption mechanism and dynamically S-Boxes have been introduced, implemented and discussed. The paper shows that the QAES development and design do not contradict the security of the AES algorithm, since all the mathematical criteria remain unchanged. The QAES symmetric encryption algorithm has been revealed depending on the integration between the AES and the QKD using two different modes, the On-line and the Off-line. The experimental results and the analysis show that the QAES produces more complicated un-breakable keys, hard to be predicted by attackers than the keys generated by the AES. However, the speed of encryption of the QAES is tiny higher (0.409 seconds) than that using the AES. The strength of the QAES lies in its ability of generating a high ratio of independence between DQS-Boxes, see **Appendix**; this ratio aids in achieving a more secured environment against most types of cryptanalysis attacks.

In the future, in order to assure the strength of the QAES, the algebraic and quantum attacks are going to be implemented, and the results are going to be analyzed.

## Appendix

This appendix contains the two randomly DQS-boxes, and relations between them.

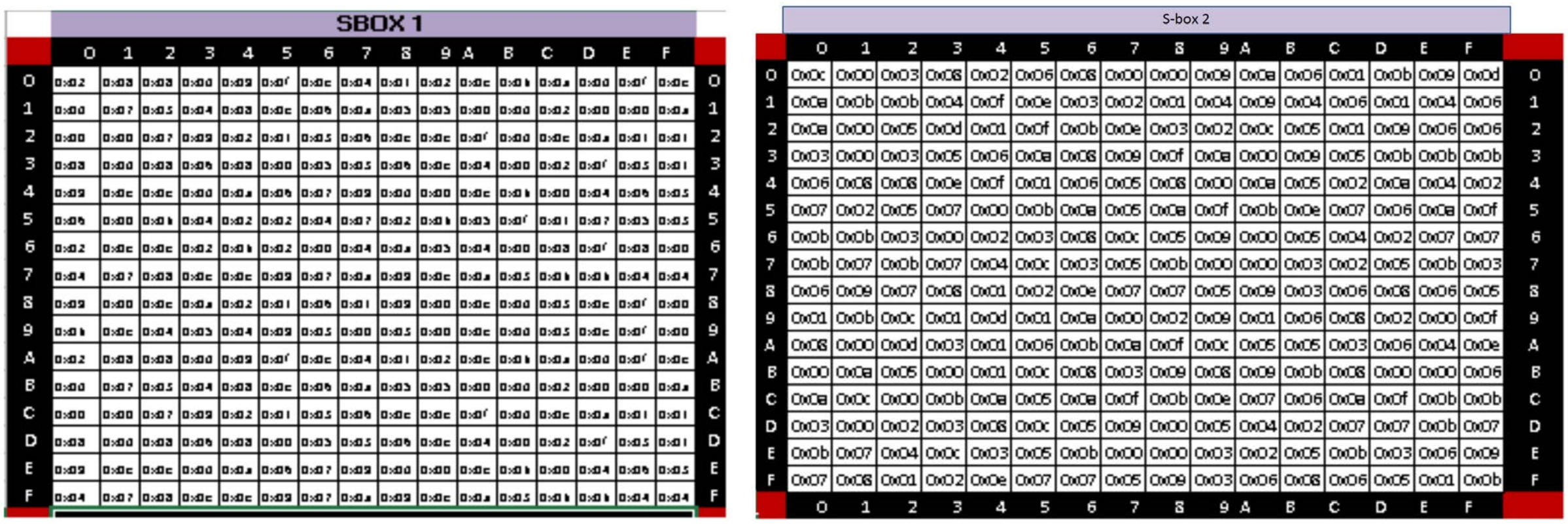

**Figure A1.** DQS-Box1 & DQS-Box2.

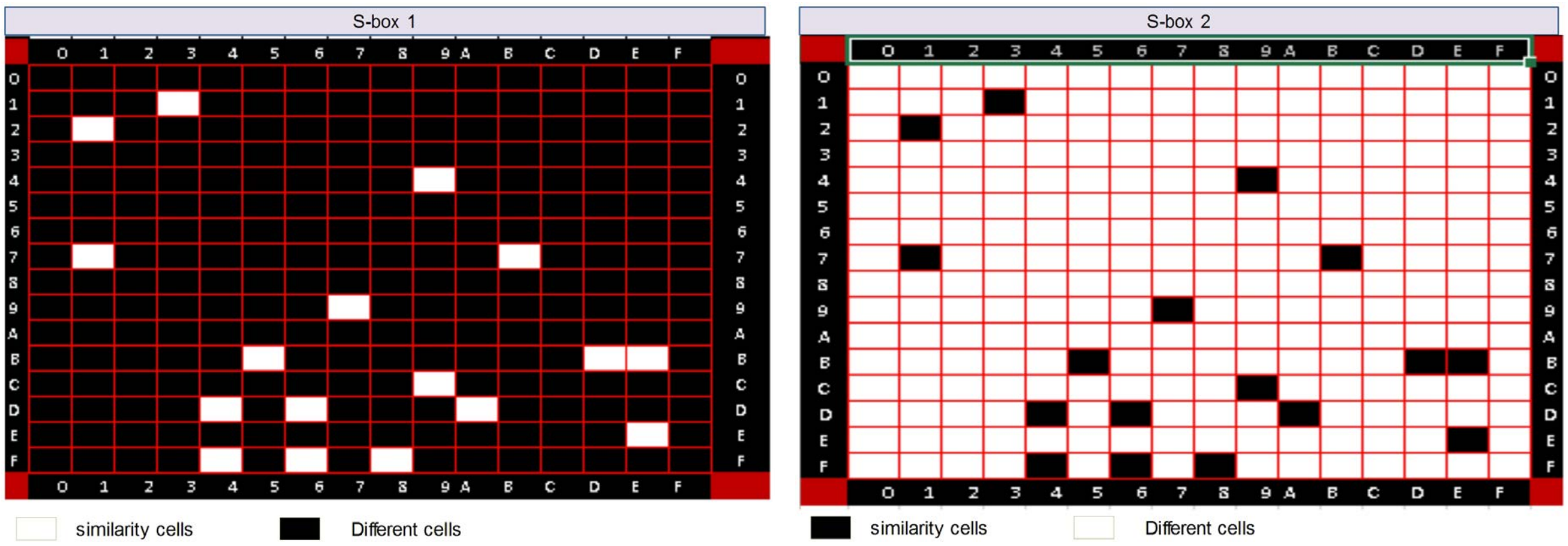

**Figure A2.** Correlation between DQS-BOX1 & DQS-BOX2.